\pdfoutput=1
\documentclass{article}
\PassOptionsToPackage{table}{xcolor}
\usepackage[T1]{fontenc}
\usepackage{iclr2027_conference,times}
\usepackage{amsmath,amssymb,amsthm}
\usepackage{graphicx}
\usepackage{float}
\usepackage{booktabs}
\usepackage{tikz}
\usetikzlibrary{arrows.meta,positioning,calc}
\usepackage{url}
\usepackage[table]{xcolor}
\usepackage{xspace}
\usepackage{hyperref}
\hypersetup{colorlinks=true,linkcolor=blue,citecolor=blue,urlcolor=blue}

\newcommand{\sysname}{\textsc{RubricCraft}\xspace}
\newcommand{\sysabbr}{RC}
\newcommand{\diffind}{cross-variant inductive synthesis}
\newcommand{\Diffind}{Cross-variant inductive synthesis}
\newcommand{\DiffInd}{Cross-variant Inductive Synthesis}

\newcommand{\rubric}{R}

\newcommand{\attempts}{N}

\definecolor{exRubric}{HTML}{1A73E8}
\definecolor{exLive}{HTML}{1E8E3E}
\definecolor{exGap}{HTML}{D93025}
\definecolor{exJudge}{HTML}{E37400}
\newcommand{\exrubric}[1]{\textcolor{exRubric}{#1}}
\newcommand{\exlive}[1]{\textcolor{exLive}{#1}}
\newcommand{\exgap}[1]{\textcolor{exGap}{#1}}
\newcommand{\exjudge}[1]{\textcolor{exJudge}{#1}}

\title{RubricCraft:\\ Where Solutions Diverge, Criteria Emerge}

\author{Hosein Hasanbeig, Priyanshu Gupta, Ashish Tiwari, Vu Le, Sumit Gulwani \\
Microsoft}

\iclrfinalcopy

\begin{document}
\maketitle
\lhead{Preprint}

\begin{abstract}
Evaluating language agents on open-ended, artifact-producing tasks is difficult: model-based graders can be inconsistent, while exhaustive programmatic checks are often impractical. Textual rubrics make evaluation criteria explicit, but writing them requires domain expertise. Rubrics generated from task descriptions alone can miss requirements and failure modes that become apparent in attempted solutions. We introduce RubricCraft, which treats rubric construction as the generation of natural-language specifications guided by candidate solutions. Given a task description and independently sampled solver trajectories, RubricCraft compares how each attempt satisfies or violates the task requirements and uses these differences to construct rubric criteria. It applies the criteria to the collected solutions and revises checks that overlook errors or penalize valid alternatives. The resulting rubrics capture partial correctness without requiring any single solution to serve as a complete reference. On SpreadsheetBench v2, against a prompt-only rubric written and applied by the same model, RubricCraft rubrics pass fewer wrong outputs (29.5\% against 42.6\%), fail none of the outputs the benchmark's own checker accepts, and rank outputs by correctness more often, while separating models as often. Repair on a single attempt leaves grading unchanged, but repair driven by attempts from three model families acts on different tasks, and an evidence audit judges most of those repairs improvements, many on tasks whose reference solution is itself wrong.
\end{abstract}

\section{Introduction}
\label{sec:intro}

Evaluating an agent that produces an artifact is harder than evaluating one that
produces an answer. A mathematics answer matches or does not; or a patch passes the
test suite or does not \citep{jimenez2024swebench}. However, a populated spreadsheet or a
drafted report has no such checker. Correctness is distributed across many
partially-satisfiable properties: whether the model used the requested headers,
avoided editing sheets it was not asked to touch, formatted the result so a human
can act on it, and refrained from asserting things the data does not support.
No single predicate decides the outcome.

The answer to this problem has converged on rubric-based grading: a list of
natural-language criteria, written by a domain expert or generated by a model,
against which a judge model scores the artifact
\citep{arora2025healthbench,starace2025paperbench,zheng2023judging}. This is a genuine improvement over
holistic judging, which is position-biased, verbosity-biased, and unstable
\citep{wang2024positionbias,panickssery2024selfpref,dubois2024lengthcontrolled}. Conditioning the judge on
explicit criteria makes the grade decomposable and auditable. However, rubrics themselves should not be treated as ground truth by construction. In practice they might fail in characteristic ways:

\begin{itemize}
\itemsep 0pt
\item \textbf{Unfair criteria.} The criterion demands something the task does not
  actually require, or requires an artifact property that is unachievable given
  the input. Every solver is penalized for a defect in the rubric.
\item \textbf{Incomplete criteria.} The rubric omits a requirement on which
  solutions genuinely differ. Two materially different artifacts receive the
  same score, and the benchmark is blind to a real capability gap.
\item \textbf{Non-discriminative criteria.} The criterion is satisfied by every
  plausible attempt (or by none). It consumes grading budget and inflates
  apparent rubric coverage while contributing no signal.
\item \textbf{Ambiguous or non-checkable criteria.} The criterion cannot be
  decided from the available evidence, so the judge's verdict is effectively a
  coin flip and flips again when the judge model is swapped.
\end{itemize}

\textbf{Our Approach: \DiffInd{}.}
A rubric cannot be simply validated by inspection, because its failure modes are
relational: a criterion is unfair only relative to what competent solvers
actually produce, and incomplete only relative to the axes on which solutions
actually differ. Both become visible under variation.
We call the resulting mechanism \textit{\diffind{}}: criteria are induced from the ways independent attempts at the same task \emph{differ}, rather than from reading the task in isolation. We therefore propose to generate \attempts{} independent attempts under varied
configurations, spanning model families, reasoning budgets, and scaffolds.
Grading them produces a grid of (criterion, attempt, verdict, evidence) that is the substrate for defect detection: a criterion failed by
attempts that a reviewer would call correct is a fairness defect; an axis on
which attempts diverge with no corresponding criterion is a coverage defect; a
criterion with constant verdict across the attempts is a discrimination defect.

Detected defects should not merely reported and ideally each should be converted into a targeted
repair of the offending criterion. Further, the repaired rubric has to be re-validated
against the same attempts. Only under this rigorous regime, the output is a rubric that has been
pressure-tested against real solver behavior rather than assumed correct.

\textbf{Evaluating the Evaluator.}
The methodological difficulty is that rubric quality cannot be measured by the
scores the rubric produces. If repair raises pass rates, the natural reading is
that grading became easier, not better; if it lowers them, that grading became
stricter, not more accurate.
We therefore organize rubric meta-evaluation around agreement with expert judgments, fidelity to independently specified checks, discrimination, and stability across judges. We call this workflow \sysname{}.
Our completed study measures descriptive model separation on fixed artifacts, agreement with the benchmark's own checks, and grading stability across repeated runs.
A larger score gap or a different ranking does not establish that a grader is more accurate.
The control is a naive rubric written by the same model from the task and source workbook alone, graded by the same artifact-inspecting judge, prompt, and tools.

We start with spreadsheets domain because they are hard to grade, and a hard domain stresses every stage of the pipeline.
Correctness is spread over many interdependent cells, e.g., in the SpreadsheetBench-v2 tasks we study \citep{zhu2026spreadsheetbench2}, the median task checks 142 target cells and must leave 6{,}374 others unchanged, and one wrong reference can propagate across sheets or be offset by another error so that the totals still come out right.
Hard is not the same as representative, however: spreadsheets are one artifact type, \sysname{} was developed on them, and its rubric generator can use the golden workbook's difference from the input.

\textbf{Contributions.}
\begin{enumerate}
\itemsep 0pt
\item \textbf{\sysname{}}, a multi-agent validation-and-repair pipeline that
  detects defects in a given rubric by \diffind{} and iteratively repairs the rubric until it
  passes validation.
\item \textbf{A controlled rubric-comparison protocol} with fixed submissions and a matched judge, distinguishing score resolution from independently established evaluation quality.
\item \textbf{An exploratory SpreadsheetBench v2 case study} comparing three graders on three solver models: grade correctness against the seed evaluator's cell checks, four rubric-construction conditions graded three times each, an audit of the repairs that diverse attempts trigger, the effect of grading evidence on stability, and two-level semantic coverage of the seed's requirements.
\item \textbf{A defect taxonomy for evaluation rubrics}: unfairness,
  incompleteness, non-discrimination, and non-checkability, each defined
  operationally with respect to a set of solution attempts.
\end{enumerate}
\section{Problem Setting and Defect Taxonomy}
\label{sec:defects}

\textbf{Setup.}
A task $t$ consists of a natural-language request and an initial context. A
rubric $\rubric_t = \{c_1,\dots,c_k\}$ is a finite set of criteria, each a
natural-language predicate over the produced artifact with an associated weight.
A solver $\pi$ maps $t$ to an artifact $a$ and a trajectory $\tau$. A judge $J$
maps $(t, a, \tau, c)$ to a verdict and supporting evidence. The benchmark score
aggregates per-criterion verdicts.

We take $\rubric_t$ as input, whether authored by an expert or generated by a
model, and ask whether it is fit to serve as the specification of correctness for $t$.

\textbf{A running example.}
The following invented task is used throughout this section; Figure~\ref{fig:running} draws it in the same colors.
A user supplies a small sales workbook and asks: \emph{``add a Total column to the Q3 sheet, and leave the other sheets unchanged.''}
A plausible rubric $\rubric_t$ has three criteria: \exrubric{(1)~the Q3 sheet gains a column labelled Total}; \exrubric{(2)~each total equals the sum of that row's monthly values}; \exjudge{(3)~no other sheet is modified}.
Now consider three attempts. Attempt~1 writes \exlive{\texttt{=SUM(B2:D2)}} in each row. Attempt~2 computes the same numbers and \exgap{pastes them as constants}. Attempt~3 writes the formulas correctly but also \exjudge{restyles a header on an untouched sheet}.

\begin{figure}[t]
\centering
\newcommand{\exsheet}[4]{%
  \draw[black!30,line width=0.35pt] ({#1-1.32},-0.30) rectangle ({#1+1.32},-0.58);
  \draw[black!30,line width=0.35pt] ({#1-0.98},-0.30) -- ({#1-0.98},-0.58);
  \node[font=\tiny\itshape,text=black!55] at ({#1-1.15},-0.44) {fx};
  \node[font=\tiny,anchor=west,inner sep=0pt] at ({#1-0.93},-0.44) {#2};
  \foreach \i/\hd/\val in {0/Jul/400,1/Aug/420,2/Sep/420} {
    \draw[black!30,line width=0.35pt,fill=black!5] ({#1-1.32+0.66*\i},-0.66) rectangle ({#1-0.66+0.66*\i},-0.94);
    \node[font=\tiny\bfseries,text=black!65] at ({#1-0.99+0.66*\i},-0.80) {\hd};
    \draw[black!30,line width=0.35pt] ({#1-1.32+0.66*\i},-0.94) rectangle ({#1-0.66+0.66*\i},-1.22);
    \node[font=\tiny,text=black!80] at ({#1-0.99+0.66*\i},-1.08) {\val};
  }
  \draw[black!30,line width=0.35pt,fill=black!5] ({#1+0.66},-0.66) rectangle ({#1+1.32},-0.94);
  \node[font=\tiny\bfseries,text=black!65] at ({#1+0.99},-0.80) {Total};
  \draw[#3,line width=0.6pt,fill=#3!12] ({#1+0.66},-0.94) rectangle ({#1+1.32},-1.22);
  \node[font=\tiny\bfseries,text=#3] at ({#1+0.99},-1.08) {1240};
  \node[draw=black!35,fill=white,font=\tiny,inner xsep=2pt,inner ysep=0.6pt,anchor=north west] at ({#1-1.32},-1.28) {Q3};
  \node[font=\tiny,inner xsep=2pt,inner ysep=0.6pt,anchor=north west,#4] at ({#1-0.92},-1.28) {Sheet2};
}
\begin{tikzpicture}[x=1cm,y=1cm,
  rowlab/.style={font=\tiny,anchor=west,text=black!80,inner xsep=0pt},
  tag/.style={font=\tiny\itshape,anchor=east},
  yes/.style={text=exLive,font=\scriptsize},
  no/.style={text=exGap,font=\scriptsize}]

\node[draw=exRubric!45,fill=exRubric!7,line width=0.55pt,rounded corners=2pt,align=left,
      inner sep=2.6pt,font=\tiny,text=black!80,text width=3.9cm,anchor=north west] at (0,-0.02)
  {\textbf{Task} $t$: add a Total column to the Q3 sheet, and leave the other sheets unchanged.};
\node[font=\tiny,text=exRubric,anchor=north west,inner xsep=0pt,text width=4.3cm,align=left] at (0,-0.98)
  {\textbf{Rubric} $\rubric_t$ (rows below) looks complete when read alone};

\node[anchor=west,inner xsep=0pt] at (4.63,-0.13) {\scriptsize\textbf{$a_1$}\ \tiny writes formulas};
\exsheet{5.95}{\textcolor{exLive}{\texttt{=SUM(B2:D2)}}}{exLive}{draw=black!35,fill=black!4,text=black!55}
\node[anchor=west,inner xsep=0pt] at (7.53,-0.13) {\scriptsize\textbf{$a_2$}\ \tiny pastes the numbers};
\exsheet{8.85}{\textcolor{exGap}{\texttt{1240}\ \ pasted value}}{exGap}{draw=black!35,fill=black!4,text=black!55}
\node[anchor=west,inner xsep=0pt] at (10.43,-0.13) {\scriptsize\textbf{$a_3$}\ \tiny formulas, restyles Sheet2};
\exsheet{11.75}{\textcolor{exLive}{\texttt{=SUM(B2:D2)}}}{exLive}{draw=exJudge,fill=exJudge!15,text=exJudge,font=\tiny\bfseries}
\node[font=\tiny,text=exJudge,anchor=west] at (11.52,-1.395) {header restyled};

\draw[black!20,line width=0.4pt] (0,-1.66) -- (13.1,-1.66);
\node[rowlab] at (0,-1.86) {\textcolor{exRubric}{\textbf{$c_1$}}\ \ the Q3 sheet gains a Total column};
\node[rowlab] at (0,-2.18) {\textcolor{exRubric}{\textbf{$c_2$}}\ \ each total equals its row's sum};
\node[rowlab,text=exJudge] at (0,-2.50) {\textbf{$c_3$}\ \ no other sheet is modified \textit{(does restyling count?)}};
\draw[exGap,dashed,line width=0.6pt,rounded corners=2pt] (-0.16,-2.68) rectangle (13.1,-3.02);
\node[rowlab,text=exGap] at (0,-2.85) {\textbf{no criterion:} the totals stay live};
\draw[black!20,line width=0.4pt] (0,-3.16) -- (13.1,-3.16);
\node[rowlab] at (0,-3.36) {\textbf{Score under $\rubric_t$}};
\node[tag,text=exGap] at (4.85,-3.36) {$a_1$ and $a_2$ tie};

\foreach \x in {5.95,8.85,11.75} {
  \node[yes] at (\x,-1.86) {$\checkmark$};
  \node[yes] at (\x,-2.18) {$\checkmark$};
}
\node[yes] at (5.95,-2.50) {$\checkmark$};
\node[yes] at (8.85,-2.50) {$\checkmark$};
\node[text=exJudge,font=\scriptsize\bfseries] at (11.75,-2.50) {?};
\node[yes] at (5.95,-2.85) {$\checkmark$};
\node[no] at (8.85,-2.85) {$\times$};
\node[yes] at (11.75,-2.85) {$\checkmark$};
\node[font=\tiny\bfseries,text=exGap] at (5.95,-3.36) {3/3};
\node[font=\tiny\bfseries,text=exGap] at (8.85,-3.36) {3/3};
\node[font=\tiny\bfseries,text=exJudge] at (11.75,-3.36) {2/3 or 3/3};
\end{tikzpicture}
\caption{\textbf{The running example: a rubric that looks complete in isolation, read against three attempts.}
All three attempts put the right totals in a new Total column, so $c_1$ and $c_2$ pass everywhere; \exlive{green} marks live formulas and passing verdicts.
\exgap{Red}: $a_1$ and $a_2$ differ on whether the totals stay live, a formula against a pasted value, yet no criterion checks it, so they tie (incompleteness).
\exjudge{Amber}: $c_3$ does not say whether restyling a header modifies Sheet2, so $a_3$'s verdict depends on the judge (non-checkability).}
\label{fig:running}
\end{figure}

Read in isolation, $\rubric_t$ looks complete. Read against these three attempts it is not.
\exgap{Attempts~1 and~2 are scored identically}, even though only one of them keeps the totals responsive to later edits, a property the requester plainly cares about and no criterion mentions.
\exjudge{Criterion~3 is stated loosely enough} that a judge may or may not count restyling as modification, so attempt~3's verdict depends on the judge rather than the artifact.
Neither problem is visible from the rubric alone; both are visible the moment the rubric is read against attempts that differ.

\subsection{Defects, defined by \diffind{}}
Let $A = \{(a_i,\tau_i)\}_{i=1}^{\attempts}$ be a set of attempts and
$V \in \{0,1\}^{k \times \attempts}$ the verdict matrix.

\textbf{Unfairness.}
Criterion $c_j$ is unfair if attempts that satisfy the task's stated intent
nonetheless fail $c_j$. Operationally, we surface criteria with high failure rates
and adjudicate them by rebuttal: the evidence for the failing verdict is
re-examined to determine whether the fault lies with the attempt or with the
criterion's phrasing or premises.

\textbf{Incompleteness.}
$\rubric_t$ is incomplete if attempts differ materially on some axis that no
criterion addresses. Detection compares attempts against each other and flags
divergences whose verdict contribution is zero under the current rubric.
The \exgap{formula-versus-constant split} between attempts~1 and~2 is exactly this case.

\textbf{Non-discrimination.}
Criterion $c_j$ is non-discriminative \emph{on the sampled attempts} if row $j$ of $V$ is constant.
This is a diagnostic flag, not sufficient evidence that the criterion is defective: a required safety or preservation constraint may legitimately be satisfied by every attempt.
Removing such criteria requires adjudication against the task, not variance alone.

\textbf{Non-checkability.}
Criterion $c_j$ is non-checkable if the judge cannot ground a verdict in evidence
from $(t,a,\tau)$. Detection uses verdict instability across repeated and
paraphrased judging.  \exjudge{Criterion~3} of the running example is the archetype.

Each defect above is defined by a pattern in $V$ or in the differences within $A$.
None is visible from $\rubric_t$ alone. This is the paper's central methodological
commitment: rubric validation requires executing the rubric against real solver
behavior, just as specification validation requires executing against real
program behavior, i.e., \diffind.

\section{\sysname{}}
\label{sec:method}

\begin{figure}[t]
\centering
\definecolor{gblue}{HTML}{1A73E8}
\definecolor{gred}{HTML}{D93025}
\definecolor{gamber}{HTML}{E37400}
\definecolor{ggreen}{HTML}{1E8E3E}
\definecolor{gpurple}{HTML}{8430CE}
\definecolor{gink}{HTML}{3C4043}
\definecolor{ggray}{HTML}{9AA0A6}
\newcommand{\robot}{\raisebox{-0.35ex}{\includegraphics[height=1.75ex]{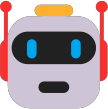}}}
\begin{tikzpicture}[
  x=1cm,y=1cm,
  font=\scriptsize,
  panel/.style={draw=ggray!55,fill=white,line width=0.6pt,rounded corners=3pt},
  chip/.style={draw=#1,fill=#1,text=white,rounded corners=2pt,
    inner xsep=2.5pt,inner ysep=1.2pt,font=\tiny\bfseries},
  card/.style={draw=gblue!45,fill=gblue!7,line width=0.55pt,rounded corners=2pt,
    align=left,inner sep=2.6pt,font=\tiny,text=gink},
  outcard/.style={draw=ggreen!55,fill=ggreen!8,line width=0.55pt,rounded corners=2pt,
    align=left,inner sep=2.6pt,font=\tiny,text=gink},
  fixcard/.style={draw=gpurple!55,fill=gpurple!7,line width=0.55pt,rounded corners=2pt,
    align=left,inner sep=2.4pt,font=\tiny,text=gink},
  tag/.style={draw=gblue!60,fill=white,rounded corners=1.5pt,inner xsep=1.6pt,inner ysep=0.6pt,
    font=\tiny\bfseries,text=gblue},
  ttl/.style={font=\scriptsize\bfseries,text=gink,anchor=west},
  note/.style={font=\tiny,text=gink,align=left},
  flow/.style={-{Stealth[length=1.7mm,width=1.3mm]},draw=gink!75,line width=0.8pt},
  loop/.style={-{Stealth[length=1.9mm,width=1.4mm]},draw=gpurple,line width=1.0pt},
  yes/.style={text=ggreen,font=\tiny},
  no/.style={text=gred,font=\tiny}
]

\draw[panel] (0,0) rectangle (4.15,-3.9);
\draw[panel] (4.5,0) rectangle (9.35,-3.9);
\draw[panel] (9.7,0) rectangle (13.3,-3.9);

\node[ttl] at (0.12,-0.27) {(a) $\attempts$ rollouts diverge};
\node[chip=gblue,anchor=east] at (4.05,-0.27) {\robot\,SOLVERS};

\node[card,text width=3.62cm,anchor=north west] at (0.12,-0.54)
  {\textbf{Task} $t$: add a Total column to the Q3\\sheet; leave other sheets unchanged.\\
   Rubric $\rubric_t$: $c_1,c_2,c_3$};

\node[card,text width=2.72cm,anchor=north west] at (0.12,-1.50)
  {$a_1$\, \texttt{=SUM(B2:D2)} per row};
\node[tag,anchor=east] at (4.05,-1.66) {model A};
\node[card,text width=2.72cm,anchor=north west] at (0.12,-2.10)
  {$a_2$\, \texttt{1240} pasted as a value};
\node[tag,anchor=east] at (4.05,-2.26) {model B};
\node[card,text width=2.72cm,anchor=north west] at (0.12,-2.70)
  {$a_3$\, formulas; restyles Sheet2};
\node[tag,anchor=east] at (4.05,-2.86) {model C};
\node[note,text=gblue,anchor=north west,text width=3.8cm] at (0.12,-3.22)
  {independent rollouts under varied models, reasoning budgets and scaffolds};

\node[ttl] at (4.62,-0.27) {(b) Grade, then compare};
\node[chip=gamber,anchor=west] at (4.64,-0.62) {\robot\,GRADER};
\node[chip=gred,anchor=west] at (6.08,-0.62) {\robot\,ANALYZER};

\foreach \x/\lab in {7.95/$a_1$, 8.50/$a_2$, 9.05/$a_3$}
  {\node[font=\tiny,text=gink] at (\x,-0.72) {\lab};}

\node[note,anchor=west,text width=3.0cm] at (4.62,-1.06) {$c_1$ Total column added};
\node[yes] at (7.95,-1.06) {$\checkmark$}; \node[yes] at (8.50,-1.06) {$\checkmark$};
\node[yes] at (9.05,-1.06) {$\checkmark$};

\node[note,anchor=west,text width=3.0cm] at (4.62,-1.46) {$c_2$ Totals numerically correct};
\node[yes] at (7.95,-1.46) {$\checkmark$}; \node[yes] at (8.50,-1.46) {$\checkmark$};
\node[yes] at (9.05,-1.46) {$\checkmark$};

\node[note,anchor=west,text width=3.0cm] at (4.62,-1.86) {$c_3$ Other sheets untouched};
\node[yes] at (7.95,-1.86) {$\checkmark$}; \node[yes] at (8.50,-1.86) {$\checkmark$};
\node[no]  at (9.05,-1.86) {$\times$};

\draw[ggray,line width=0.5pt] (9.20,-0.94) -- (9.28,-0.94) -- (9.28,-1.58) -- (9.20,-1.58);
\node[note,text=ggray,anchor=north west,text width=4.5cm] at (4.62,-2.02)
  {$c_1,c_2$ agree on every rollout: no signal (merge)};

\node[font=\tiny\bfseries,text=gred,anchor=south west] at (4.60,-2.80) {\DiffInd};
\draw[gred,dashed,line width=0.7pt,rounded corners=2pt]
  (4.58,-2.82) rectangle (9.30,-3.22);
\node[note,text=gred,anchor=west,text width=3.0cm] at (4.68,-3.02)
  {\textbf{no criterion:} totals stay live};
\node[yes] at (7.95,-3.02) {$\checkmark$}; \node[no] at (8.50,-3.02) {$\times$};
\node[yes] at (9.05,-3.02) {$\checkmark$};
\node[note,text=gred,anchor=north west,text width=4.6cm] at (4.60,-3.28)
  {$a_1$ and $a_2$ differ here, yet every criterion\\scores them the same: induce one};

\node[ttl] at (9.82,-0.27) {(c) Repair};
\node[chip=gpurple,anchor=east] at (13.2,-0.27) {\robot\,FIXER};

\node[fixcard,text width=3.18cm,anchor=north west] at (9.82,-0.56)
  {\textbf{Induce} $c_4$: totals stay live};
\node[fixcard,text width=3.18cm,anchor=north west] at (9.82,-1.02)
  {\textbf{Sharpen} $c_3$: any cell edit counts};
\node[fixcard,text width=3.18cm,anchor=north west] at (9.82,-1.48)
  {\textbf{Merge} $c_1,c_2$ into one};

\node[draw=gred!70,fill=gred!6,rounded corners=6pt,font=\tiny,text=gink,align=center,
      inner sep=2.4pt,text width=2.9cm] (accept) at (11.42,-2.36)
  {\textbf{Analyzer re-checks $\rubric'$}\\on the \emph{same} rollouts: accepted?};
\draw[flow] (11.42,-1.93) -- (accept.north);

\node[outcard,text width=3.18cm,anchor=north west] (frozen) at (9.82,-3.22)
  {\textbf{yes:} freeze $\rubric^{*}$, content-hashed};
\draw[flow] (accept.south) -- (11.42,-3.22);

\draw[flow] (4.15,-1.9) -- (4.5,-1.9);
\draw[flow] (9.35,-1.2) -- (9.7,-1.2);

\draw[loop] (accept.east) -- (13.55,-2.36) -- (13.55,0.40) -- (6.93,0.40) -- (6.93,0);
\node[font=\tiny\bfseries,text=gpurple,anchor=south] at (13.34,-2.36) {no};
\node[anchor=south,font=\tiny,text=gpurple,align=center] at (9.80,0.44)
  {\raisebox{-0.3ex}{\includegraphics[height=1.6ex]{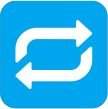}}\ \textbf{FEEDBACK LOOP:}
   re-grade and re-analyze on the same cached rollouts, $\leq k$ rounds};

\draw[ggray!45,line width=0.5pt] (0,-4.18) -- (13.55,-4.18);
\node[anchor=north west,inner sep=0pt] at (0.02,-4.30) {\includegraphics[height=0.42cm]{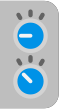}};
\node[font=\tiny\bfseries,text=gink,anchor=north west,align=left] at (0.36,-4.30) {Control\\knobs};
\node[note,anchor=north west,text width=3.0cm] at (1.30,-4.28)
  {\textcolor{gblue}{\textbf{(a) rollouts}} $\attempts$, model mix,\\reasoning budget, temperature};
\node[note,anchor=north west,text width=3.0cm] at (4.40,-4.28)
  {\textcolor{gblue}{\textbf{(a) rubric origin}} reuse the\\benchmark seed, or induce fresh};
\node[note,anchor=north west,text width=3.0cm] at (7.50,-4.28)
  {\textcolor{gred}{\textbf{(b) criterion budget}} count,\\weight total, granularity};
\node[note,anchor=north west,text width=3.0cm] at (10.60,-4.28)
  {\textcolor{gpurple}{\textbf{(c) acceptance}} majority rule,\\effect tolerance, rounds $k$};

\end{tikzpicture}
\caption{\textbf{\sysname{} is an agentic workflow that induces criteria from the ways rollouts differ.}
Running example: a rubric that looks complete in isolation is incomplete once it is read against attempts.
\textbf{(a)}~Solver agents produce $\attempts$ independent rollouts under varied configurations.
\textbf{(b)}~A grader agent fills the verdict matrix with cited workbook evidence, and an analyzer agent compares the rollouts.
Rows $c_1$ and $c_2$ agree on every rollout, so they separate nothing; $a_1$ and $a_2$ differ on whether the totals stay live, yet \emph{no row covers that axis}.
Finding such an axis by comparing rollouts, and turning it into a criterion, is \emph{differential induction} (dashed box).
\textbf{(c)}~A fixer agent induces the missing criterion, sharpens the ambiguous one, and merges the redundant pair.
The \emph{feedback loop} then re-grades and re-analyzes the repaired rubric on the same cached rollouts, so no score change can come from a newly sampled solution; a rejected rubric returns to (b) for another round, and an accepted one is frozen.
The strip lists the control knobs the user sets at each stage.}
\label{fig:pipeline}
\end{figure}
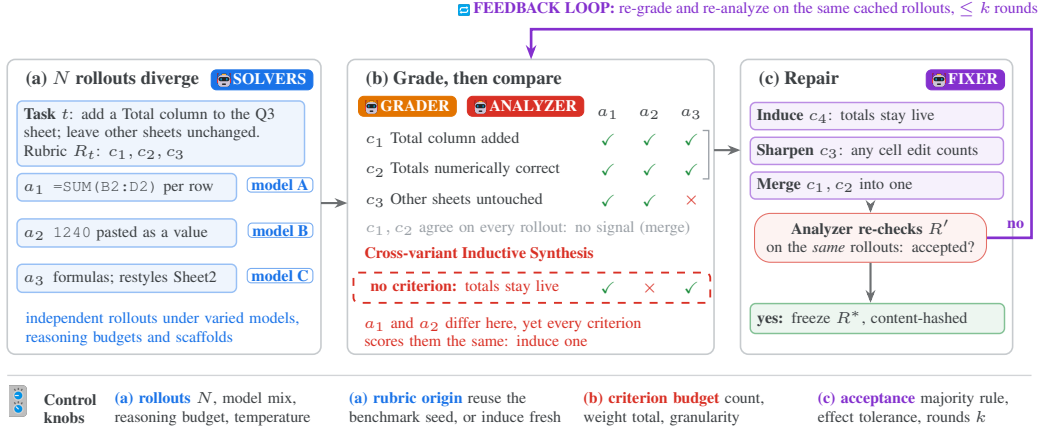

\sysname{} is organized as four cooperating roles operating in a bounded loop (Figure~\ref{fig:pipeline}).

\textbf{Solver: constructing the attempts.}
\sysname{} generates \attempts{} attempts under a configuration schedule chosen to
\emph{widen behavioral support} rather than to maximize quality: varied model
families, reasoning budgets, scaffolds, and decoding temperature.
A set of uniformly strong attempts is nearly as uninformative as a uniformly weak
one, because in both cases $V$ collapses toward constant rows and the defect
signals that \diffind{} depends on vanish.

\textbf{Grader: producing verdicts with evidence.}
Each attempt is graded against each criterion, and the judge must emit the evidence supporting its verdict, not only the verdict: a failure rate alone cannot distinguish an unfair criterion from a hard one.

\textbf{Analyzer: defect detection.}
The analyzer consumes $(V, \text{evidence}, A)$ and emits a typed defect report per criterion, applying the detectors of Section~\ref{sec:defects}, plus a coverage report of divergence axes with no corresponding criterion.
Each defect carries an estimated effect on the score, so defects are ranked by how much they would change grading rather than merely flagged.
A rubric is accepted when fairness, coverage, and conciseness each hold for a configurable majority of attempts and any residual defect falls below a configurable effect tolerance; attempts whose solver or analyzer call failed are excluded rather than counted as passing.

\textbf{Fixer: targeted repair and re-validation.}
Each defect becomes a scoped edit: rephrase an unfair criterion, add a criterion for an uncovered axis, merge or delete non-discriminative criteria, or replace a non-checkable criterion with a grounded one.
Repairs keep a flat weighted checklist with a fixed point total, so revisions change content rather than scale.
The loop re-validates the repaired rubric and halts when validation accepts it, a stage errors, the repair returns an unchanged rubric, or the round budget is exhausted; each repair prompt carries the previous rounds' change summaries, and on exhaustion the pipeline keeps the candidate with the smallest combined fairness and coverage effect rather than the last one attempted.
When the attempts are cached, re-validation regrades the same artifacts, so a score change across rounds reflects the rubric rather than a newly sampled solution.

\textbf{Freezing and cost.}
An accepted rubric is serialized and content-hashed against the task, so its cost, dominated by the \attempts{} solves and gradings rather than by rubric editing, is paid once; each later evaluation run is a single grading pass.

\section{Meta-Evaluation Protocol}
\label{sec:metaeval}

A rubric's quality cannot be read off the scores it produces, so we evaluate the grader along four axes.
As in the CogEval protocol for evaluating LLM capabilities \citep{momennejad2023cogeval}, comparisons rest on a control condition, repeated runs, and statistical tests rather than on single scores: a matched naive rubric, regrading of the same outputs, and paired bootstrap intervals.
Let $s_{tmg}\in[0,100]$ be the reported percentage score for task $t$, model $m$, and grader $g$.
For \sysabbr{} and naive, $s_{tmg}=25u_{tmg}$, where $u_{tmg}\in[0,4]$ is the raw judge grade; for seed, $s_{tmg}=100y_{tmg}$, where $y_{tmg}\in\{0,1\}$ is the benchmark's own task outcome.
An \sysabbr{} score of 85\% therefore means 85\% of the maximum grade, not an 85\% task-success rate; score differences are percentage points (pp).

\textbf{Agreement.}
Do grades agree with human judgments of which outputs are right?
Assuming that the benchmark's golden workbooks were constructed and checked by experts, the seed evaluator's cell comparison stands in for a human verdict on every target cell.
Let $c_{tm}\in[0,1]$ be the seed's modified-cell accuracy, the fraction of target cells whose value (or formula, or font color) matches the golden workbook.
We call an output \emph{wrong} when $c_{tm}<0.8$ and report the false-pass rate $\Pr[u_{tmg}\geq 3.5\mid c_{tm}<0.8]$ and the false-fail rate $\Pr[u_{tmg}<3\mid y_{tm}=1]$.
The proxy inherits any errors in the golden workbooks, of which Section~\ref{sec:results} documents several, and it sees only the compared cells, so disagreement with it bounds error rather than measuring it.

\textbf{Fidelity.}
Does the grade move with satisfaction of the task's requirements, crediting accepted alternatives and withholding credit where a requirement is unmet?
We report how often a grader fails near-complete work ($c_{tm}\geq 0.95$) and work whose only differences lie in cells the task did not ask to change, and how closely its criteria match the seed's requirements, extracted as text (Appendix~\ref{app:semcov}).

\textbf{Discrimination.}
Do graders resolve differences between fixed solutions?
With $\bar s_{mg}$ the mean of $s_{tmg}$ over a common task set, we report the model-mean spread $\Delta_g=\max_m\bar s_{mg}-\min_m\bar s_{mg}$ and the share of tasks whose best and worst model scores differ by at least 12.5 pp; the seed outcome differs by either 0 or 100 pp, so its share is the same for any positive threshold.
A common display scale does not make graded credit and binary success the same construct, and more separation is not by itself better grading.

\textbf{Stability.}
Does the rubric produce the same verdict when the same artifact is graded again?
We grade the same outputs three times under one judge configuration and report the verdict-flip rate and within-task standard deviation.

\textbf{Baselines.}
The study compares \textbf{\sysabbr{}} (the finalized \sysname{} rubric), \textbf{naive} (a prompt-only rubric from the same model), and \textbf{seed}, SpreadsheetBench v2's own programmatic cell comparator.
Seed is neither human grading nor the initial rubric $R_0$: for 297 of 321 tasks the benchmark ships no text rubric, so \sysname{} generates $R_0$ itself.
We also grade the unrepaired $R_0$ and a rubric repaired on attempts from three model families (Appendix~\ref{app:conditions}).

\section{Experiments}
\label{sec:experiments}

\textbf{Benchmark and cohort.}
We use SpreadsheetBench v2 \citep{zhu2026spreadsheetbench2}, the successor of SpreadsheetBench \citep{ma2024spreadsheetbench}, pinned to dataset revision \texttt{5a2215ed} and harness commit \texttt{5c160265} and released under the MIT license.
\sysname{} produced a finalized rubric for 302 tasks, 278 of them nonvisual; the 24 visualization tasks are excluded because the seed evaluator compares cells.
The primary cohort is the 259 nonvisual tasks (80 debugging, 89 financial-model, and 90 template) on which all three solvers submitted an output and every grader returned a score.

\textbf{Solvers and graders.}
GPT-5.6 Sol, GPT-6 Astra, and Opus~5 solve every task through mini-swe-agent 2.4.6 (the names identify the tested development endpoints), and their outputs are frozen before grading, so every grader scores the same workbook bytes.
Naive is a rubric written by a separate GPT-5.6 Sol session from the task text and a read-only view of the source workbook, with the instruction ``write a grading rubric for the following task'' and no rubric-writing guidance, examples, or golden workbook.
Both rubrics are applied by the same GPT-5.6 Sol artifact-inspecting judge, with the same harness, prompt, workbook-inspection tools, reasoning setting, and budget; the judge returns a report and a grade on $[0,4]$.
Seed is the benchmark's published comparator, run through an archived local wrapper: it passes an output when every target cell matches the golden workbook and at least 99.8\% of preserved cells are unchanged, comparing values on 242 cohort tasks, formulas on nine, and font colors on eight.

\textbf{Rubric construction.}
\sysabbr{} was built from GPT-5.6 Sol attempt. A rubric was accepted when a majority of non-errored attempts passed validation and any residual defect moved the grade by less than 0.8 of 4 points.
\sysname{}'s generator uses an engineered checklist prompt and receives the source-to-golden workbook difference as a sample solution, so the \sysabbr{}-versus-naive contrast measures the whole construction pipeline.
Over the full 321-task run, 244 initial rubrics passed without repair and 58 were repaired, averaging 13.9 model requests and 575K tokens per task and 4.95 hours of wall time in total at concurrency five (Appendix~\ref{app:convergence}); each repair round of this run re-sampled the attempt rather than regrading the cached one.
A second construction, loop + diversity, repairs the same $R_0$ on the three solvers' frozen outputs, which it regrades in every round.

\textbf{Grading evidence and statistics.}
The submitted workbooks were saved without computed values, so every grader reads an Excel-recalculated copy of each submission, bound to it by the submission's hash.
We call an output \emph{wrong} when fewer than 80\% of its target cells match and \emph{near-complete} when at least 95\% do; a rubric \emph{passes} an output at a grade of at least $3.5/4$ and \emph{fails} it below $3/4$.
Intervals are 95\% paired task-bootstrap intervals.

\section{Results}
\label{sec:results}

\begin{figure}[t]
\centering
\includegraphics[width=\linewidth]{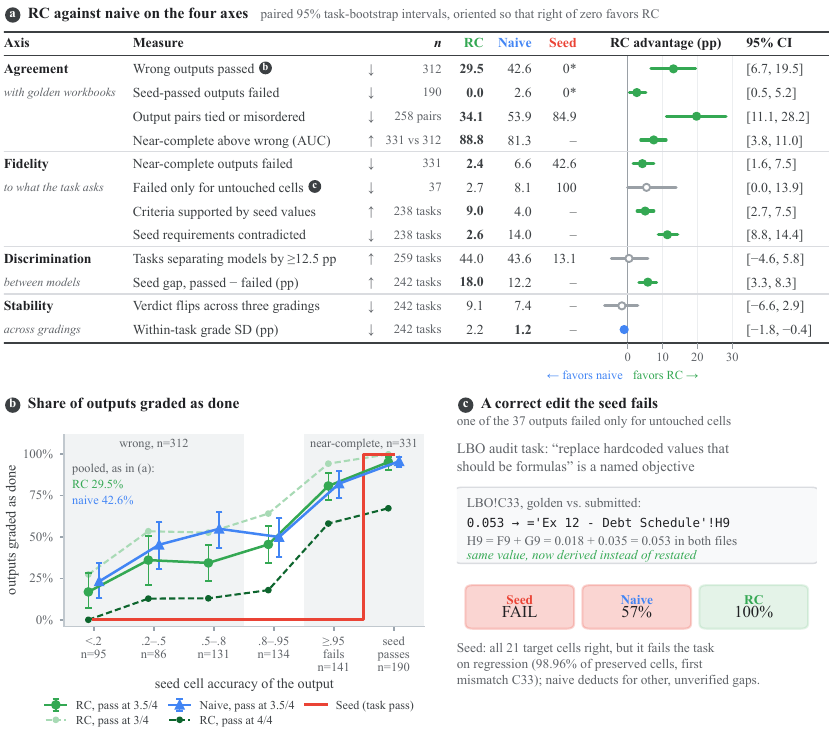}
\caption{\textbf{\sysabbr{} against the naive rubric on the four axes, with the two views behind the marked rows.}
(a)~Values are percentages unless marked pp, and $n$ counts outputs unless marked; the seed-gap and stability rows grade GPT-6 Astra's outputs three times.
Intervals are 95\% paired task-bootstrap intervals for \sysabbr{}'s advantage over naive, signed so that right of zero favors \sysabbr{} (green: excludes zero in its favor; blue: in naive's; gray: includes zero); bold marks the better value where the interval excludes zero.
AUC is the probability that a near-complete output is graded above a wrong one. $^\ast$Zero by construction.
(b)~Share of the 777 outputs graded as done, by the seed's cell accuracy; the row marked~b pools the three wrong bands at $3.5/4$, and the seed's task bit (red) passes nothing below its all-cells threshold.
Dashed lines move \sysabbr{}'s threshold to $3/4$ and $4/4$.
(c)~One of the 37 outputs in the row marked~c, verified from the stored cells (Appendix~\ref{app:example-0302}).}
\label{fig:results}
\end{figure}

\textbf{Diverse attempts expose defects that a single attempt misses.}
Repair on the three solvers' frozen outputs and repair on one attempt changed 29 and 30 of the 259 cohort rubrics, only 9 in common: 20 of the diversity repairs fall on tasks whose $R_0$ the single attempt accepted.
Where diversity repaired, the rubric covers more of the seed's required fixes than $R_0$ (48.6\% against 38.5\% on 27 tasks, $[+3.6,+17.7]$ pp), whereas single-attempt repairs show no such gain (35.9\% against 39.1\%, $[-8.4,+0.5]$ pp).
An audit of 19 of the 20 diversity-only repairs, in which reviewer agents checked each repair against the cells where solver outputs and the golden workbook disagree, judged 11 improvements, 4 regressions, and 4 neutral, and found the golden reference itself wrong on 11: when most of three solver families disagree with a rubric, the fault often lies in the rubric's reference.
The same repairs contradict more of the seed's expected values (10.4\% against 0.2\%) and pass more of Astra's wrong outputs on these tasks (61\% against 11\% of 19), partly because they override wrong references and partly through regressions, such as accepting a mistake all three solvers share.
Since they were validated on the very outputs graded here and audited by agents, these repairs are evidence for the mechanism rather than a held-out gain (Appendix~\ref{app:conditions}).

\textbf{Agreement.}
Figure~\ref{fig:results}a compares \sysabbr{} with the naive rubric on the four axes of Section~\ref{sec:metaeval}: eight of its twelve intervals favor \sysabbr{}, three include zero, and one favors naive.
\sysabbr{} passes 29.5\% of the 312 wrong outputs against 42.6\% for naive, fails none of the 190 outputs that the seed passes (naive 2.6\%), and ties or reverses 34.1\% of the 258 same-task pairs whose cell accuracies differ by at least 5 points (naive 53.9\%, seed bit 84.9\%).
Figure~\ref{fig:results}b shows where the first row comes from: both rubrics' pass rates rise with the seed's cell accuracy, \sysabbr{}'s stays below naive's in the three wrong bands, and the seed bit stays at zero until every target cell is right.
The gap does not hinge on the pass threshold (AUC 88.8\% against 81.3\%; at $4/4$, 9.0\% against 12.8\% of wrong outputs passed), on the cut that defines wrong, or on the nine formula-comparison tasks (Appendix~\ref{app:agreement}).
Because the audit above found golden workbooks wrong, some counted false passes are corrections of the reference, and the rates bound disagreement rather than measure error.

\textbf{Fidelity.}
The seed bit fails 42.6\% of the 331 near-complete outputs and all 37 whose only differences lie in cells the task did not ask to change; \sysabbr{} fails 2.4\% and 2.7\% of them, naive 6.6\% and 8.1\%.
Figure~\ref{fig:results}c is one of the 37: in a leveraged-buyout audit that asks for ``hardcoded values that should be formulas'' to be fixed, the submission links \texttt{LBO!C33} to a debt-schedule cell with the same value, $0.053$, and the seed fails the task on preserved-cell mismatches, the first of them this cell, while \sysabbr{} awards full marks under a criterion that accepts equivalent live links and naive awards $2.3/4$ for other gaps we have not verified.
\sysabbr{} criteria are also fully supported by the seed's value-level requirements twice as often as naive criteria (9.0\% against 4.0\%) and contradict a fifth as many (2.6\% against 14.0\%), while naive rubrics, 4.9 times longer, cover more of the seed's fixes (57\% against 46\%).

\textbf{Discrimination and stability.}
The seed bit gives the three models the same outcome on 225 of the 259 tasks, whereas both rubrics separate them by at least 12.5 pp on about 44\% of tasks; \sysabbr{} ties its separations to correctness, grading Astra's seed-passed outputs 18.0 pp above its seed-failed ones against 12.2 pp for naive.
Naive's larger model-mean spread (10.21 against 4.89 pp) comes almost entirely from debugging tasks, on which it grades Astra 27 and 30 pp above the other two models.
In each of three gradings \sysabbr{} passes 29.5--35.2\% of Astra's 105 wrong outputs and naive 45.7--48.6\%; verdict flips do not distinguish the rubrics, and \sysabbr{}'s grades vary more within a task, the one measure on which naive is better.
Both rates depend on computed values: graded from the uncached submissions, \sysabbr{} flipped 16.5\% of verdicts and passed 43--52\% of the wrong outputs.
The unrepaired $R_0$ matches \sysabbr{} on these measures (seed gap 19.0 pp, 32.7\% false passes, 6.2\% flips), so at $\attempts=1$ the gains come from \sysname{}'s generator and its golden-difference evidence rather than from repair (Appendix~\ref{app:conditions}).

\section{Related Work}
\label{sec:related}

\textbf{LLM judges and their reliability.}
Using a strong model as a grader became standard with MT-Bench and Chatbot Arena \citep{zheng2023judging}, G-Eval's form-filling scores \citep{liu2023geval}, and evidence that model ratings track expert ratings of open-ended text \citep{chiang2023llmeval}; \citet{gu2025surveyjudge} survey the area.
The same judges are position-biased \citep{wang2024positionbias}, prefer their own generations \citep{panickssery2024selfpref}, and reward length unless it is controlled \citep{dubois2024lengthcontrolled}, and on hard, objectively checkable response pairs strong judges barely beat chance \citep{tan2025judgebench}; RewardBench meta-evaluates reward models in the same spirit \citep{lambert2024rewardbench}.
ALLURE audits an LLM evaluator against annotated data and iteratively adds the instances where it deviates significantly to its in-context examples \citep{hasanbeig2023allure}.
This line of work audits the judge.
We hold the judge fixed and audit the specification it is given, and the evidence result of Section~\ref{sec:results} adds a third source of grading error: what the judge is shown, consistent with the finding that LLMs do not reliably detect or correct subtle distortions in table representations \citep{dutta2026tabular}.

\textbf{Rubrics and checklists as evaluation specifications.}
Conditioning the judge on explicit criteria is the main remedy.
FLASK decomposes scoring into skill-level criteria \citep{ye2024flask}, Prometheus trains an open evaluator on score rubrics \citep{kim2024prometheus}, and BiGGen Bench writes instance-specific criteria for every item \citep{kim2025biggen}.
InFoBench scores the fraction of decomposed requirements a response meets \citep{qin2024infobench}, and TICK has a model generate the checklist from the instruction \citep{cook2024tick}.
Artifact-level benchmarks now ship rubrics as their ground truth: physician-written criteria in HealthBench \citep{arora2025healthbench}, author-co-developed hierarchical rubrics for research replication in PaperBench \citep{starace2025paperbench}, expert-graded occupational deliverables in GDPval \citep{patwardhan2025gdpval}, and hierarchical requirements checked by an agentic judge in DevAI \citep{zhuge2024agentjudge}, whose artifact-inspecting judge resembles ours.
In all of these the rubric is written once from the task, by an expert or a model, and then treated as correct; \sysname{} instead tests and repairs it against attempts.
The closest precedent is EvalGen \citep{shankar2024validators}, which reports \emph{criteria drift}: users need criteria to grade outputs, grading outputs changes the criteria, and some criteria exist only relative to the outputs observed.
\Diffind{} automates that observation, using divergence among attempts in place of a user's grades.

\textbf{Benchmark validity and grader error.}
Measurement quality is increasingly treated as a property of the benchmark itself.
Test sets carry label errors that can reorder models \citep{northcutt2021labelerrors}, most benchmarks omit significance testing and replication support \citep{reuel2024betterbench}, and agentic benchmarks misestimate performance through flawed task setups and reward designs, such as insufficient tests or empty responses counted as success \citep{zhu2025abc}.
The code-generation analogue of our false passes is EvalPlus, where 80 times more tests caught previously undetected wrong programs and changed model rankings \citep{liu2023evalplus}.
Item response theory treats evaluation items as more or less informative and discriminating \citep{rodriguez2021informative,polo2024tinybenchmarks}; our discrimination axis asks the same of graders, and the non-discrimination defect is its criterion-level form.
Spreadsheet benchmarks grade by comparing output cells with a reference workbook \citep{li2023sheetcopilot,ma2024spreadsheetbench,zhu2026spreadsheetbench2}, the kind of binary comparator we use as the seed; the golden workbooks it compares against are themselves sometimes wrong (Section~\ref{sec:results}).
Conversational data-analysis benchmarks, whose goals are often under-specified, instead grade answers with an LLM evaluator checked against human annotations and rate the dialogue against fixed rubrics \citep{dutta2026condabench}.

\textbf{Specification inference and testing the tests.}
The idea that a specification can be recovered or checked by observing a
population of implementations rather than by inspecting the specification alone
originates in program analysis. Dynamic invariant detection infers likely
invariants from many executions \citep{ernst2007daikon}; specification mining
recovers API contracts from traces \citep{ammons2002specmining}; and
differential testing treats disagreement among independent implementations of the same specification as evidence of a fault or an under-determined specification \citep{mckeeman1998differential,yang2011csmith}, the premise of N-version programming \citep{avizienis1985nversion}. The same idea recovers task structure in reinforcement learning, where an automaton synthesized from exploration traces exposes the unknown sequence of high-level objectives a task requires \citep{hasanbeig2021deepsynth,hasanbeig2024symbolic}. \sysname{} transposes this to a setting where the
``executions'' are solver attempts, the ``invariants'' are natural-language
criteria, and agreement must be judged semantically rather than by equality.
The transposition removes the sound checker that makes the program-analysis
version tractable, which is precisely why our meta-evaluation
(Section~\ref{sec:metaeval}) relies on external gold signals rather than internal
consistency.
Software testing also evaluates its own evaluators: mutation analysis scores a test suite by the seeded faults it detects \citep{demillo1978hints}, mutant detection correlates with real-fault detection \citep{just2014mutants}, and the oracle problem names the difficulty of deciding correct behavior at all \citep{barr2015oracle}.
A rubric is a test oracle written in natural language, and our false-pass and false-fail rates against the seed's cells play the role of a mutation score.

\textbf{Evaluation signals in post-training.}
Post-training on checkable tasks rewards programmatic verifiers \citep{lambert2024tulu,guo2025deepseek}, and group-based methods derive advantages from comparisons among rollouts \citep{shao2024deepseekmath}.
Beyond language models, reinforcement learning also turns formal task specifications into rewards: a temporal-logic requirement is compiled into an automaton that shapes the reward, with guarantees that the learned policy satisfies the requirement with maximal probability \citep{hasanbeig2019temporal,hasanbeig2023certified}.
Rubric-derived scores extend such rewards to open-ended tasks: Rubrics as Rewards optimizes against instance-specific rubrics in medicine and science \citep{gunjal2025rar}, and checklist feedback outperforms reward models for instruction following \citep{viswanathan2025checklists}.
Optimizing an imperfect proxy degrades the true objective \citep{gao2023overoptimization}, so once a rubric becomes a reward its defects become the policy's target.
Our work is upstream of this line: rather than proposing a reward, it asks whether the
criteria such rewards are computed from are fair, complete, and discriminative in
the first place. The observation that distilling per-attempt comparisons into a
persistent symbolic rubric makes an otherwise transient signal reusable is a
motivation for this work rather than a claim we evaluate here.

\section{Conclusion}
\label{sec:conclusion}

Rubrics are evaluation instruments, and their quality can be measured.
On SpreadsheetBench v2, \sysname{} rubrics improve on a naive rubric from the same model and judge in agreement, fidelity, and discrimination: they pass fewer wrong outputs (29.5\% against 42.6\%), fail less near-complete work (2.4\% against 6.6\%), contradict a fifth as many of the benchmark's expected values, and tie model separation to correctness (a seed gap of 18.0 against 12.2 pp), with verdicts as reproducible as naive's when the judge reads computed values.
With one attempt the gains come from construction; with three model families, repair acts on different tasks and an audit judges most of those repairs improvements.

The four axes also separate effects that a single score conflates.
A naive rubric already turns the benchmark's pass/fail bit into graded scores that separate the three models by at least 12.5 pp on about 44\% of tasks, against 13\% for the seed bit, but separation is informative only when it tracks correctness: most of naive's larger model-mean spread comes from grading one model far above the other two on debugging tasks while passing more outputs that the golden workbooks reject.
Evidence matters as much as criteria: recalculating each workbook before grading roughly halved the \sysname{} rubrics' verdict flips and removed a third of their false passes, so how an artifact is rendered for the judge is part of the grading instrument.
Divergence among attempts pinpoints where a rubric needs attention: when independent solver families agree against a reference, the reference deserves a second look before the solvers are penalized.

These gains are robust and cheap to obtain.
The advantage in agreement does not hinge on the pass threshold (AUC 88.8\% against 81.3\%), on the cut that defines a wrong output, or on the formula-comparison tasks, and it holds in each of three repeated gradings, in none of which \sysabbr{} fails a seed-passed output.
The loop converges quickly, with 80\% of initial rubrics accepted unrepaired and 99\% within two repairs, and without inflating scores; because an accepted rubric is frozen, its one-time cost of 13.9 model requests per task leaves every later evaluation a single grading pass.
Diverse attempts extend its reach: repair driven by three model families revised rubrics on 20 tasks that a single attempt had accepted, raised coverage of the benchmark's required fixes from 38.5\% to 48.6\% where it acted, and surfaced golden references that an audit judged wrong on 11 of 19 tasks.

\bibliographystyle{iclr2027_conference}
\bibliography{references}

\clearpage
\appendix
\section{Repair-Loop Dynamics}
\label{app:convergence}

\begin{figure}[H]
\centering
\includegraphics[width=\linewidth]{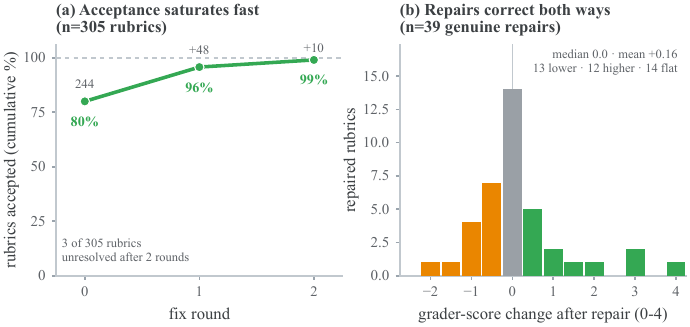}
\caption{\textbf{The repair loop converges fast and does not inflate scores.}
(a)~Cumulative share of the 305 gradable rubrics accepted by fix round on the full 321-task run (16 infrastructure-skipped tasks excluded): 80\% pass unrepaired, 96\% within one repair, and 99\% within two, with three left unresolved; the third permitted repair round was never needed.
(b)~For the 39 repairs carrying a clean analyzer verdict at both ends and a genuinely flagged initial defect, the change in the grader's score from the first to the final round; because each round re-sampled the attempt, the change mixes rubric and attempt differences.
The distribution centers on zero (median $0$; 13 lower, 12 higher, 14 unchanged; mean $+0.16$ of four points), so repair does not systematically raise scores.
A further 16 of the 60 multi-round cases were retries after a transient non-JSON analyzer response rather than substantive repairs, and are excluded from~(b).}
\label{fig:convergence}
\end{figure}

The repair loop converges quickly and does not inflate scores (Figure~\ref{fig:convergence}): acceptance saturates within two repair rounds, and on the repairs with a genuine analyzer-flagged defect the grader's score moves in both directions between the first and the final round, with a median change of zero.
In this run every repair round re-sampled the solver attempt rather than regrading the cached one, so the change mixes the rubric's revision with a new attempt, and each repaired rubric was accepted on a fresh attempt.
Within the 259-task cohort, 30 rubrics were revised; Appendix~\ref{app:conditions} shows that at $\attempts=1$ these revisions do not measurably change grades on a held-fixed solver's outputs.

\section{Rubric Conditions and Grading Evidence}
\label{app:conditions}

\begin{figure}[H]
\centering
\includegraphics[width=\linewidth]{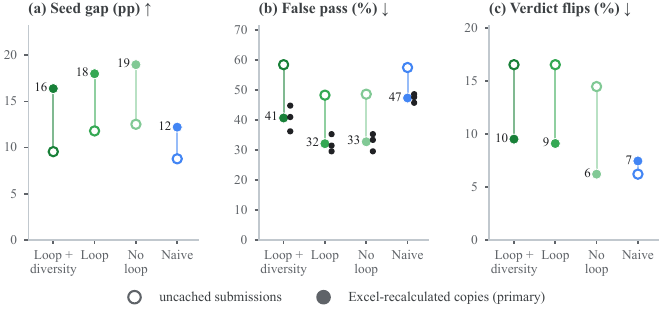}
\caption{\textbf{Four rubric conditions, each grading the same 242 GPT-6 Astra outputs three times, from the uncached submissions (hollow) and from their Excel-recalculated copies (filled).}
(a)~Seed gap: mean grade on seed-passed minus seed-failed outputs.
(b)~False pass: share of the 105 outputs with fewer than 80\% of target cells right that are graded as done; dots are the three recalculated runs.
(c)~Verdict flips: share of outputs whose pass/fail verdict changes across the three gradings.
Recalculated evidence roughly halves the \sysname{} conditions' flips and removes a third of their false passes.}
\label{fig:conditions}
\end{figure}

We grade GPT-6 Astra's frozen outputs on 242 tasks three times under each of four rubric conditions, from the recalculated copies and, for comparison, from the uncached submissions (Figure~\ref{fig:conditions}, Table~\ref{tab:conditions}).

\textbf{Reproducibility.}
The three run means of every condition lie within 0.9 pp of one another, and no condition fails any of the 61 seed-passed outputs in any run.
Averaged over runs, \sysabbr{}'s false-pass rate is 15.2 pp lower than naive's ($[-24.1,-6.9]$), its seed gap 5.8 pp larger ($[+3.3,+8.3]$), and its verdicts match the seed bit 9.8 pp more often (62.4\% against 52.6\%, $[+4.5,+15.0]$).

\textbf{Stable grading requires computed values.}
Graded from the uncached submissions, the \sysname{} conditions flipped verdict on 14.5--16.5\% of outputs and passed 43--62\% of the wrong outputs across runs.
The judge noticed the missing values in high- and low-scoring runs alike; what varied was whether it then evaluated the formulas or read stale values from the workbook difference.
In \texttt{template:13\_06}, one run took the input's cached gross profit as the submitted result and scored the corrected model 1.6/4, while the other two evaluated the formulas and gave 4/4.
In \texttt{template:01\_08}, GPT-5.6's bond model books the whole \$15{,}900 final payment as coupon, so its extinguishment gain is \$15{,}825 where the golden workbook has \$825; from the uncached file the \sysabbr{} judge read the formulas as isolating the principal and gave $4/4$, and from the recalculated copy the same rubric gives $2.7/4$ and names both errors.
Recalculation cut the \sysname{} conditions' flips by 7.0--8.3 pp and their false passes by 15.9--17.8 pp, with every interval excluding zero; naive's flips did not change ($+1.2$ pp, $[-2.9,+5.0]$).

\begin{table}[H]
\caption{Rubric conditions graded three times from recalculated copies of the same 242 GPT-6 Astra outputs (percent of maximum grade; ranges over the three runs).
The seed gap is the mean grade on seed-passed minus seed-failed outputs, false pass uses the 105 outputs with fewer than 80\% of target cells right, and flips are outputs whose pass/fail verdict changes across the three runs.
No condition fails any of the 61 seed-passed outputs in any run.}
\label{tab:conditions}
\centering
\small
\setlength{\tabcolsep}{3.2pt}
\begin{tabular}{@{}llrrrr@{}}
\toprule
Condition & Construction & Mean (runs) & Seed gap & False pass & Flips \\
\midrule
Loop + diversity & repaired, 3-model attempts & 86.0 (85.5--86.3) & 16.4 & 40.6\% & 9.5\% \\
Loop (\sysabbr{}) & repaired, Sol attempt & 85.0 (84.9--85.2) & 18.0 & 32.1\% & 9.1\% \\
No loop & first-round $R_0$ & 84.2 (84.0--84.4) & 19.0 & 32.7\% & 6.2\% \\
Naive & prompt-only & 89.6 (89.5--89.7) & 12.2 & 47.3\% & 7.4\% \\
\bottomrule
\end{tabular}
\end{table}

\section{Additional Agreement Results}
\label{app:agreement}

\begin{figure}[tbp]
\centering
\includegraphics[width=\linewidth]{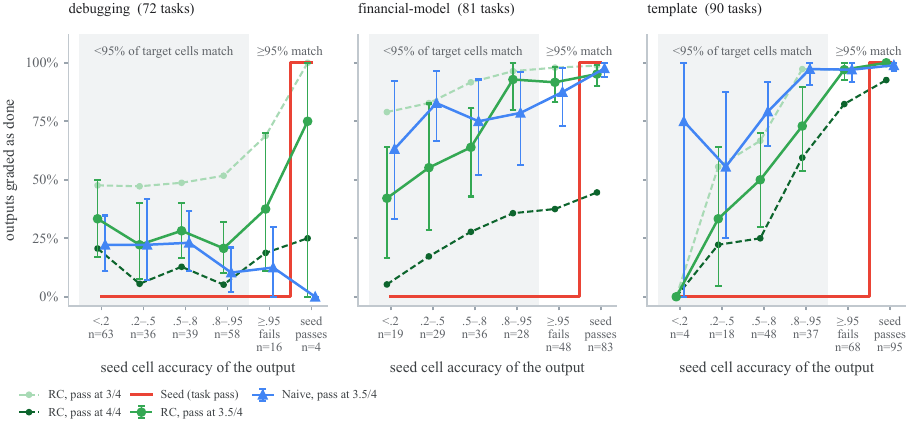}
\caption{\textbf{Figure~\ref{fig:results}b split by task family.}
Bins with few outputs (for example the four seed-passed debugging outputs) have wide intervals and should not be read individually.
On financial-model and template tasks, \sysabbr{} at $3.5/4$ passes fewer low-accuracy outputs than naive; it matches naive on near-complete template outputs but passes fewer near-complete financial-model outputs that the seed fails (75\% against 91\% of 53).
On debugging, where almost no output fixes every seeded error, both rubrics pass few low-accuracy outputs, and \sysabbr{} passes more of the seed-failed outputs with at least 95\% of target cells right (40\% against 10\% of 20).
Moving \sysabbr{}'s threshold to $4/4$ lowers its curve most on financial-model tasks, where many outputs earn high but not perfect grades.}
\label{fig:correctness-family}
\end{figure}

Table~\ref{tab:errors} gives the full agreement and fidelity counts behind Figure~\ref{fig:results}a, including the strictest pass threshold.
When wrong instead means fewer than half of the target cells right, \sysabbr{} passes 26.0\% of those outputs and naive 33.7\% ($[-14.7,-1.1]$ pp).
Table~\ref{tab:formula-sensitivity} repeats the four main comparisons without the nine formula-comparison tasks, whose prepared copies are not recalculated; every interval still excludes zero.
Figure~\ref{fig:correctness-family} splits Figure~\ref{fig:results}b by task family.

\begin{table}[tbp]
\caption{Grading errors against the golden workbooks on the 259-task cohort (all three models' outputs).
A rubric passes an output at a grade of at least $3.5/4$ and fails it below $3/4$; the seed passes and fails by its own rule.
Intervals are 95\% paired task-bootstrap intervals for \sysabbr{} minus naive, in pp.
$^\ast$Zero by construction: the reference is the seed's own cells.}
\label{tab:errors}
\centering
\small
\setlength{\tabcolsep}{3.5pt}
\begin{tabular}{@{}lrrrrr@{}}
\toprule
Error & $n$ & Seed & \sysabbr{} & Naive & \sysabbr{}$-$naive \\
\midrule
Passes an output with $<80\%$ of target cells right & 312 & 0\%$^\ast$ & 29.5\% & 42.6\% & $[-19.5,-6.7]$ \\
Fails an output the seed passes & 190 & 0\%$^\ast$ & 0.0\% & 2.6\% & $[-5.2,-0.5]$ \\
Fails an output with $\geq 95\%$ of target cells right & 331 & 42.6\% & 2.4\% & 6.6\% & $[-7.5,-1.6]$ \\
Fails an output whose only errors are in preserved cells & 37 & 100\% & 2.7\% & 8.1\% & $[-13.9,0.0]$ \\
Ties or reverses a pair $\geq 5$ pts apart in cell accuracy & 258 & 84.9\% & 34.1\% & 53.9\% & $[-28.2,-11.1]$ \\
\midrule
\multicolumn{6}{@{}l}{\emph{Strictest rubric setting, pass only at $4/4$} (seed row: its task bit)} \\
Passes an output with $<80\%$ of target cells right ($\downarrow$) & 312 & 0\%$^\ast$ & 9.0\% & 12.8\% & -- \\
Passes an output with $\geq 95\%$ of target cells right ($\uparrow$) & 331 & 57.4\% & 63.4\% & 49.8\% & -- \\
\bottomrule
\end{tabular}
\end{table}

\begin{table}[tbp]
\caption{The main agreement and fidelity comparisons without the nine formula-comparison tasks (250 tasks), against the full cohort (259 tasks).
Intervals are 95\% paired task-bootstrap intervals for \sysabbr{} minus naive, in pp.}
\label{tab:formula-sensitivity}
\centering
\small
\setlength{\tabcolsep}{3.5pt}
\begin{tabular}{@{}lrrrrrr@{}}
\toprule
& \multicolumn{3}{c}{All 259 tasks} & \multicolumn{3}{c}{Without formula mode (250)} \\
\cmidrule(lr){2-4}\cmidrule(l){5-7}
Measure & \sysabbr{} & Naive & Interval & \sysabbr{} & Naive & Interval \\
\midrule
Wrong outputs passed & 29.5\% & 42.6\% & $[-19.5,-6.7]$ & 29.5\% & 44.4\% & $[-21.7,-8.2]$ \\
Seed-passed outputs failed & 0.0\% & 2.6\% & $[-5.2,-0.5]$ & 0.0\% & 2.1\% & $[-4.7,-0.5]$ \\
Near-complete outputs failed & 2.4\% & 6.6\% & $[-7.5,-1.6]$ & 2.1\% & 5.5\% & $[-6.8,-0.6]$ \\
Pairs tied or misordered & 34.1\% & 53.9\% & $[-28.2,-11.1]$ & 33.3\% & 55.4\% & $[-30.8,-13.8]$ \\
\bottomrule
\end{tabular}
\end{table}

\section{Semantic Coverage Protocol for SpreadsheetBench v2}
\label{app:semcov}
\begin{figure}[tbp]
\centering
\includegraphics[width=\linewidth]{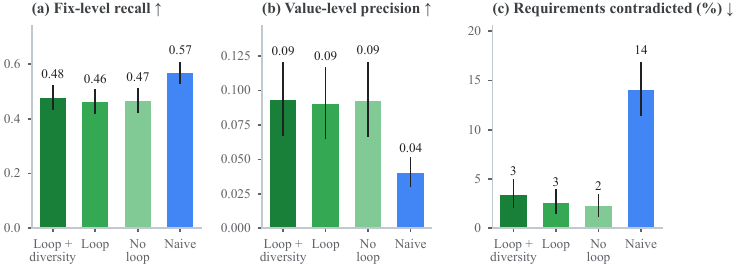}
\caption{\textbf{\sysname{} rubrics agree with the seed's requirements; naive rubrics enumerate more of them.}
Semantic coverage on the 238 tasks where every rubric set completed (task means with 95\% task-bootstrap intervals; only full matches count).
(a)~Share of fix-level requirements (what must be fixed, and where) that a rubric fully covers.
(b)~Share of rubric criteria fully supported by value-level requirements (each fix with up to two key expected values).
(c)~Share of value-level requirements that a rubric explicitly contradicts.}
\label{fig:semcov}
\end{figure}

The protocol behind the coverage rows of Figure~\ref{fig:results}a reuses a published criterion-overlap implementation unchanged for the matching prompt, response validation, and metrics, and adapts only what SpreadsheetBench v2 requires.
It was run on 243 tasks, all inside the primary cohort, and 238 of them completed for every rubric set (Figure~\ref{fig:semcov}).

\begin{itemize}
\itemsep 0pt
\item \textbf{Reference documents.} Per task: the instruction (context only), the published evaluator source, and a check sheet generated with the evaluator's own cell classification, listing each compared cell's input and expected value by sheet row, the mode flags, and the count of preserved cells. Adjacent cells of a row share one line; numbers are shown to seven significant digits, far finer than the evaluator's 1\% tolerance. All 243 tasks fit the 240{,}000-character input bound without truncation.
\item \textbf{Extraction.} The implementation's extraction instructions are used verbatim, followed by one SpreadsheetBench paragraph that sets the granularity. Both levels state one requirement per fix (a corrected error or restored structure in repair tasks, one built section in build tasks) and the preservation check once. Fix-level requirements carry no expected values, tolerances, or cell lists and cite only the check sheet's row headers, so the matcher never sees values; value-level requirements name at most two key expected values, and the comparison rule (tolerance and mode) is stated once as its own requirement. Every requirement must quote a verifier document exactly. The fix level extracted all 243 tasks (median 8 requirements, maximum 52); the value level extracted 241, with two failing quote validation after three attempts.
\item \textbf{Generated criteria.} Checklist rubrics split at checklist items; naive rubrics, which are free-form, split at scored headings (one scored section per criterion) without discarding text; three naive rubrics with no headings are excluded.
\item \textbf{Matching.} Both directions in batches of at most 24 targets against all candidates, with labels full, partial, none, contradicted, and uncertain; identical requests are answered once, so rubric conditions with identical text share results. The matching prompt's provenance sentence is replaced by a SpreadsheetBench one.
\item \textbf{Choice of granularity.} Four extraction styles were compared on the same 12-task pilot before the full runs. References that list every compared cell, its expected value, and the tolerance gave recall 0.15--0.17 for \sysname{} and 0.085 for naive, because a rubric of four to six criteria spot-checks values rather than enumerating them. The implementation's instructions alone produced evaluator-mechanics requirements (for example, that all 1{,}598 modified cells match) and recall 0.10--0.14 for \sysname{} and 0.17 for naive. The two levels reported were the remaining styles. Value-free references cannot fully support criteria that state values, so fix-level precision is near zero by construction (0.005--0.016 on the full run) and is not reported.
\end{itemize}
The extracted reference is model-generated, not human-validated, and inherits the seed's golden-workbook errors.

\section{Worked Example}
\label{app:example}

\subsection{A representation mismatch that fails a correct edit}
\label{app:example-0302}

This is the worked form of the case summarized in Figure~\ref{fig:results}c.
It is included because every step is verifiable from the stored artifacts by direct cell inspection, without relying on any judge's opinion.

\textbf{Task.}
Task \texttt{debugging:03\_02} supplies a leveraged-buyout workbook and the instruction to
``audit and fix this file thoroughly,'' listing the admissible error classes explicitly:
formula logic errors, \emph{hardcoded values that should be formulas}, broken references,
data-validation issues, calculation errors, inconsistent conventions, and best-practice violations.
Replacing a hardcoded value with an equivalent formula is therefore not an incidental liberty; it is one of the named objectives of the task.

\textbf{What the model did.}
Among other edits, the submission replaced the constant in \texttt{LBO!C33} with a reference to the debt schedule, as shown in Table~\ref{tab:example-cells}.
The referenced cell is itself live, so the submitted workbook computes the same rate from its two components instead of restating it.

\begin{table}[h]
\caption{Cell-level comparison for \texttt{debugging:03\_02}, read directly from the stored golden and submitted workbooks.
Only the audited cell differs, and it differs in representation rather than value: the two component cells are identical in both files, so both workbooks evaluate the rate to $0.053$.}
\label{tab:example-cells}
\centering
\small
\begin{tabular}{@{}lll@{}}
\toprule
Cell & Golden & Submitted \\
\midrule
\texttt{LBO!C33} & \texttt{0.053} (constant) & \texttt{='Ex 12 - Debt Schedule'!H9} \\
\texttt{Ex 12 - Debt Schedule!H9} & \texttt{=F9+G9} & \texttt{=F9+G9} \\
\texttt{Ex 12 - Debt Schedule!F9} & \texttt{0.018} & \texttt{0.018} \\
\texttt{Ex 12 - Debt Schedule!G9} & \texttt{0.035} & \texttt{0.035} \\
\bottomrule
\end{tabular}
\end{table}

Since $0.018+0.035=0.053$, the submitted workbook and the golden workbook agree on the value of the audited cell.
They differ only in whether that value is stated or derived, which is precisely the distinction the task asked the model to act on.

\textbf{What each evaluator concluded.}
The seed evaluator shipped with the benchmark runs this task in formula-comparison mode.
It marked all 21 modification cells correct, giving 100\% modification accuracy, and failed the task on regression: 4{,}390 of 4{,}436 preserved cells matched, for 98.96\% regression accuracy.
The first reported regression error is \texttt{LBO!C33}, with expected \texttt{0.053} against submitted \texttt{='Ex 12 - Debt Schedule'!H9}.
Because our wrapper promotes regression accuracy to a pass only at $0.998$ or above, 98.96\% is recorded as a task failure.
The failure is therefore produced entirely by cells the model was \emph{not} asked to change, under a comparator that treats a live reference to an equal value as a mismatch.

The \sysname{} rubric awarded full marks.
Its highest-weighted criterion (3.0 of 10 points) asks the model to
``replace duplicated hardcodes with formulas linked to authoritative cells'' and states that
``equivalent live-link formulas are acceptable,'' and the grader scored the submission $4/4$, or 100\%.
The naive rubric scored it $2.3/4$, or 57.5\%: it credited the debt-schedule repairs but deducted chiefly for a missing interest-deduction cap schedule and for forecast roll-forwards left untouched, deductions we have not verified.
Because the seed runs this task in formula-comparison mode, its prepared copy is not recalculated; the equivalence above is read from constants and needs no computed values.

\end{document}